\documentclass[12pt]{article}
 \usepackage{amssymb} 
\usepackage{amsmath,amscd,graphics}

 \def\be#1\ee{\begin{equation}#1\end{equation}}
 \def\bln#1\eln{\begin{array}{l}#1\displaystyle\end{array}}

 \def\bma#1\ema{{\allowdisplaybreaks\begin{align}#1\end{align}}}
 \def\nnm{\notag}

 \def\bgr#1\egr{{\allowdisplaybreaks\begin{gather}#1\end{gather}}}
 \def\ef#1{(\ref{#1})}
 \def\qef#1{$(\ref{#1})$}

\def\R{\mathbb{R}}
\def\eps{\varepsilon}

       \newtheorem{lemma}{\bf Lemma}[section]
       \newtheorem{theorem}[lemma]{\bf Theorem}

       \newtheorem{remark}[lemma]{\bf Remark}

\begin{document}

\title{\textbf{\LARGE  Semiclassical and relaxation
limits of bipolar quantum hydrodynamic model$^*$}
 \footnotetext{*Corresponding author: Hai-Liang Li}}

\author{\textbf{Guojing Zhang$^{1)}$,~~Hai-Liang Li$^{2)}$,
~~Kaijun Zhang$^{1)}$}\\[2mm]
{\it\small $^{1)}$School of Mathematics and Statistics,
 Northeast Normal University}\\
  {\it\small Changchun 130024, P.R.China }\\
{\it\small $^{2)}$Department of Mathematics, Capital Normal
University~~~~~~~}\\
{\it\small Beijing 100080, P. R. China }\\
{\it\small  email:~zhanggj112@nenu.edu.cn $($G.Z$)$,
hailiang.li.math@gmail.com $($H.L$)$}
\\
{\it\small  zhangkj201@nenu.edu.cn $($K.Z$)$}}\date{}

\maketitle \vspace{-0.5cm}
\renewcommand{\thefootnote}{\fnsymbol{footnote}}
 \begin{abstract}
\noindent\normalsize{The global in-time semiclassical and
relaxation limits of the bipolar quantum hydrodynamic model for
semiconductors are investigated in $R^3$. We prove that the unique
strong solution converges globally in time to the strong solution
of classical bipolar hydrodynamical equation in the process of
semiclassical limit and to that of the classical Drift-Diffusion
system under the combined relaxation and semiclassical limits.}
\end{abstract}

\textbf{Key words}: \begin{minipage}[t]{120mm} Quantum
hydrodynamics; semiclassical limit; relaxation limit.
\end{minipage}
\bigskip


\section{Introduction}
Recently, the quantum hydrodynamic(QHD) model for semiconductors is
derived and studied in the modelings and simulations of
semiconductor devices (like MOSFET and RTD) in ultra-small size (say
nano-size), where the effects of quantum mechanics, such as particle
tunneling through potential barriers and  built-up in quantum well,
are taken into granted and dominate  the transportation of electron
and/or hole under the self-consistent electric field.

The basic observation concerning the quantum hydrodynamics is that
the energy density consists of one additional new quantum correction
term of the order $O(\hbar)$ introduced first by
Wigner~\cite{Wigner32} in 1932, and that the stress tensor contains
also an additional quantum correction part~\cite{AncTie87,AncIaf89}
related to the quantum Bohm potential (or internal
self-potential)~\cite{Bohm52}
\be
Q(\rho) =
-\frac{\hbar^2}{2m}\frac{\Delta\sqrt{\rho}}{\sqrt{\rho}},\label{disp-2}
\ee
with observable $\rho>0$ the density, $m$ mass, and $\hbar$ the
Planck constant. The quantum potential $Q$ was introduced by de
Broglie and explored by Bohm to make a hidden variable theory and is
responsible for producing the quantum behavior, so that all quantum
features are related to its special properties.  Such possible
relation was also implied in the original idea initialized by
Madelung~\cite{Madelung27} in 1927 to derive quantum fluid-type
equations, in terms of Madelung's transformation applied to wave
function of Schr\"odinger equation of pure state. In fact, based on
this idea, one is able to derive quantum fluid type equations from
the (nonlinear) Schr\"odinger equation of
pure-state~\cite{GM97,J2001}.
\par

The moment method is employed recently to derive quantum
hydrodynamic equations for semiconductor device at nano-size based
on the Wigner-Boltzmann (or quantum Liouville)
equation~\cite{MRS1990}
\be
W_t+\xi\cdot \nabla_xW +\frac{q}{m}\mathbb{P}[\Phi]W=[W_t]_c
\label{WB}
\ee
where $W=W(x,\xi,t),\ (x,\xi,t)\in\R^3\times\R^3\times\R_+$ is the
distribution function, and $\mathbb{P}$ the pseudo-differential
operator defined by
\[
\mathbb{P}[\Phi]W=\frac{im}{(2\pi)^{N}}\int\!\!\int
\frac{\Phi(x+\frac{\hbar}{2m}\eta)-\Phi(x-\frac{\hbar}{2m}\eta)}{\hbar}
 e^{i\eta\cdot (\xi-\xi')}W(x,\xi',t)d\eta d\xi'.
\]
The electrostatic potential $\Phi=\Phi(x,t)$ is self-consistent
through Poisson equation
\[
 \lambda_0\,  \Delta \Phi= q(\int Wd\xi -\mathcal{C}),
\]
with $\lambda_0>0$ the permitivity characteristic of device, $q$ the
elementary charge, and ${\cal C}={\cal C}(x)>0$ the given doping
profile \cite{{MRS1990}}, and $[W_t]_c$ refers to the collision
term. In fact, applying moment method to the Wigner-Boltzmann
equation \ef{WB} near the ``momentum-shifted quantum
Maxwellian"~\cite{Wigner32} together with appropriate closure
assumption~\cite{G409-427,GMR94}, one can obtain the quantum
hydrodynamic equation~\cite{G409-427}. For more derivation and
related topics on the modelling of quantum models, one refers
to~\cite{MRS1990,GM97,G409-427} and the references therein.

In the present paper, we consider the bipolar quantum hydrodynamic
model of semiconductors (for carriers of two type)
\bgr
  \partial _t\rho_i+\nabla \cdot(\rho_iu_i)=0,\label{1.1}\\
  \partial_t(\rho_i u_i)+\nabla\cdot(\rho_iu_i \otimes u_i)
 +\nabla P_i(\rho_i)=q_i\rho_iE
 +\frac{\varepsilon^2}{2}\rho_i
  \nabla(\frac{\triangle\sqrt{\rho_i}}{\sqrt{\rho_i}})
 -\frac{\rho_iu_i}{\tau_i},\label{1.2}\\
 \lambda^2\nabla\cdot E=\rho_a-\rho_b-\mathcal{C},~
 \nabla \times E =0,~E(x)\rightarrow 0,~|x|\rightarrow +\infty,\label{1.3}
 \egr
where $(x,t)\in R^3\times R^+$ and the index $i=a,b$ and
$q_a=1,~q_b=-1$. The observable $\rho_a>0,\rho_b>0,u_a,u_b$ and $E$
are the densities, velocities and electric field, respectively.
$P_a(.),P_b(.)$ are the pressure-density functions. The parameters
$\varepsilon>0$, $\tau_a=\tau_b=\tau>0$, and $\lambda>0$ are the
scaled Planck constant, momentum relaxation time, and Debye length
respectively. $\mathcal{C}=\mathcal{C}(x)$ is doping profile.
\par

In the real  simulations of semiconductor devices, the size of the
device is rather small (in nanosize, for instance). This in turn
makes the scaled parameters $\tau,\varepsilon,\lambda$ rather
smaller due to different situations under
consideration~\cite{Sze69,MRS1990}. In general, the scaled
parameters $\eps,\tau,\lambda$ are expressed as
$$
 \varepsilon^2= \frac{\hbar^2}{2m\kappa_B T_0L^2}, \quad
 \lambda^2= \frac{\lambda_0\kappa T_0}{Nq^2L^2}, \quad
 {\tau}^2= \frac{\kappa_B T_0\tau_0^2}{mL^2}
$$
where we recall that the physical parameters are the elementary
charge $q$, the Boltzmann constant $k_B$, the elective electron mass
$m$, the reduced Planck constant $\hbar$, the permittivity
$\lambda_0$, the ambient temperature $T_0$, and the characteristic
device length $L$  and density $N$.  The typical values of the
parameters for semiconductors are given in \cite{MRS1990}.
Therefore, one of the both mathematically and physically important
problems is to justify the asymptotic approximation (or behavior) of
the macroscopic observable of the quantum hydrodynamical model
subject to the small parameters mentioned above.

In the present paper, we investigate the asymptotical analysis with
respect to the scaled small parameters of bipolar time-dependent
quantum hydrodynamical model. To begin with, let us present a
complete description about the small-scale asymptotics of the QHD
model. We first consider the semiclassical limit. Let
$\varepsilon\rightarrow0$ formally in (\ref{1.1})-(\ref{1.3}),  we
get the well-known bipolar hydrodynamic (HD)
model~\cite{GHL2003,GJ663-685}
 \bgr
\partial_t\rho_i+\nabla\cdot(\rho_iu_i)=0,\label{1.5}\\
\partial_t(\rho_iu_i)+\nabla\cdot(\rho_iu_i \otimes u_i)+ \nabla
P_i(\rho_i)=q_i\rho_iE -\frac{\rho_iu_i}{\tau_i},\label{1.6}\\
\lambda^2\nabla\cdot E=\rho_a-\rho_b-\mathcal{C},~\nabla \times E
=0,~E(x)\rightarrow 0,~|x|\rightarrow +\infty.\label{1.7}
 \egr
This limiting process shows the semiclassical approximation of
bipolar quantum hydrodynamical model in terms of bipolar
hydrodynamical model for small Planck constant, and describes the
relation from quantum mechanics to the classical Newtonian
mechanics.

The semiclassical limits of the stationary unipolar quantum
hydrodynamical model (carrier of one type) are well studied
recently. In one dimensional bounded domain, the semiclassical limit
of the thermal equilibrium solutions \cite{bgms} and the isentropic
subsonic solutions~\cite{GJ773-800} are analyzed respectively due to
different boundary conditions. This limit is also investigated for a
stationary unipolar viscous quantum hydrodynamical system
\cite{GJ183-203} for a special class of viscosity in one-dimensional
interval subject to the boundary condition of density and quantum
Fermi potential, where the communication between vanishing viscosity
and semiclassical limit is also investigated in subsonic regime. For
bipolar stationary quantum hydrodynamical model, the semiclassical
limits are investigated in multi-dimensional bounded domain for
isothermal solutions in thermal equilibrium
state~\cite{{U69-88},{Lz2005}}, by recovering the minimizer of
limiting functional of a quantized energy functional corresponding
the original system, and  in multi-dimensional unbounded domain for
stationary isentropic system \cite{ZZ2005}. A rigorous analysis is
also made for the bipolar viscous quantum hydrodynamical
system~\cite{{Lz2005}}.

However, all those analysis for stationary problems can not apply to
the time-dependent case because that unlike the case of stationary
problem, the maximum principle usually does not apply to the
time-dependent case and it is not clear how to derive enough the
a-priori estimates with respect to time (derivatives) so as to pass
into the semiclassical limit. Although such process of semiclassical
limit has been investigated recently for nonlinear Schr\"{o}dinger
equation\cite{{LL2001},{DLT}} for potential flow in terms of
Friedrich-Kato-Lax's theory and is concerned with the finite (short)
time theory, the frame work does not apply here to general
multi-dimensional rotational (non-potential) flow and is not fit for
global in-time theory. We should do the semiclassical limit for QHD
model in a different way in order to present the global in-time
semiclassical limit for general non-potential flow.
\par

Next, we turn to the analysis of relaxation limit. To this end, let
us introduce the diffusion scaling as \cite{JL2005,MN129-145}
\be
x\rightarrow x,~~~~t\rightarrow
\frac{t}{\tau},~~~~(\rho_i^\tau,u_i^\tau,E^\tau)(x,t)
=(\rho_i,\frac{u_i}{\tau},E)(x,\frac{t}{\tau}). \label{scaling}
\ee
 Then (\ref{1.1})-(\ref{1.3}) can be rewritten as
\be
\partial _t\rho_i^\tau+\nabla
\cdot (\rho_i^\tau u_i^\tau)=0,\label{1.8} \ee\be
\tau^2\partial_t(\rho_i^\tau u_i^\tau)
 +\tau^2\nabla\cdot(\rho_i^\tau u_i^\tau \otimes u_i^\tau)
 +\nabla P_i(\rho_i^\tau)=q_i\rho_i^\tau E^\tau
 +\mbox{$\frac{\varepsilon^2}{2}
   \rho_i^\tau\nabla(
   \frac{\Delta\sqrt{\rho_i^\tau}}{\sqrt{\rho_i^\tau}})$}
 -\rho_i^\tau u_i^\tau,\label{1.9} \ee\be \lambda^2\nabla\cdot
E^\tau=\rho_a^\tau-\rho_b^\tau-\mathcal{C}(x),~\nabla \times E^\tau
=0,~E^\tau(x)\rightarrow 0,~|x|\rightarrow +\infty,
  \label{1.10}
 \ee
Also formally, let $\tau\rightarrow0$ in (\ref{1.8})-(\ref{1.10}),
the quantum drift-diffusion(QDD) model is obtained
 \be
\partial_t\rho_i+\nabla[q_i\rho_iE-\nabla P_i(\rho_i)
 +\mbox{$\frac{\varepsilon^2}{2}\rho_i
  \nabla(\frac{\triangle\sqrt{\rho_i}}{\sqrt{\rho_i}})$}]=0,\label{1.11}
\ee
 \be
 \lambda^2\nabla\cdot E=\rho_a-\rho_b-\mathcal{C}(x),
 ~\nabla \times E=0,~E(x)\rightarrow 0,~|x|\rightarrow +\infty, \label{1.12}
 \ee
This limiting process provides a singular approximation of quantum
hydrodynamical model via parabolic quantum Drift-Diffusion model for
small momentum relaxation time. Note that although there are many
results obtained for classical hydrodynamic model
\cite{{GJ663-685},{MN129-145},{JP1007},{JP385-396}}, few is known
for the relaxation limit for the quantum hydrodynamical model due to
the less of enough information to control the nonlinear third order
dispersion term. Although the relaxation limit of the stationary
solutions are investigated  in one-dimensional bounded domain for
unipolar case \cite{GJ773-800}, and in multi-dimensional bounded
domain for bipolar case \cite{Lz2005}, like the situation of
semiclassical analysis, all these studies seems not enough in the
resolution of the time-dependent problems. Note that, the singular
relaxation time limit presented above is not mathematically
rigorous, the first rigorous analysis result about relaxation time
limit of QHD model has been obtained recently in \cite{JL2005},
where the QHD system is proven to be approximated by a quantum
Drift-Diffusion model(QDD), a nonlinear parabolic equation, for
small relaxation time. However, this analysis depends strongly on
the effects of the nonlinear dispersion. That is, the scaled Planck
constant is required to be fixed in order to help getting enough
control to pass into the relaxation limit. This analysis is
therefore not enough to prove the relaxation limit for possibly
arbitrary small Planck constant $\varepsilon$. Thus, it is natural
for us to consider the relaxation limit of quantum hydrodynamical
model for any small Planck constant $\varepsilon$ and furthermore
the combined relaxation and semiclassical limit. In fact, we can
show in the present paper that one can derive the following limiting
drift-diffusion (DD) model
\be
  \partial_t\rho_i
 +\nabla[q_i\rho_iE-\nabla P_i(\rho_i)]=0, \label{1.13}
\ee \be \lambda^2\nabla\cdot
E=\rho_a-\rho_b-\mathcal{C}(x),~\nabla \times E
=0,~E(x)\rightarrow 0,~|x|\rightarrow +\infty, \label{1.14}
 \ee
by setting $\tau\rightarrow0$ and $\varepsilon\rightarrow0$ in
(\ref{1.8})-(\ref{1.10}) for strong solutions. Note here that
although we only deal with the combined relaxation and semiclassical
limits for the quantum hydrodynamical model \qef{1.8}--\qef{1.10},
we claim that the analysis made here does not require any
(communication) restriction between $\varepsilon$ and $\tau$. That
is, one can fix any of the two parameters $\varepsilon$ and $\tau$
and let the other tend to zero.\par\bigskip

We shall also mention the asymptotical analysis about the zero-Debye
length limit for QHD model.  This process is quite well understood
for both stationary problems \cite{{U69-88},{{GJ773-800}}} for one
and multi-dimension bounded domain respectively and the
time-dependent problem for multi-dimension~\cite{LL195-212}. We omit
the corresponding analysis here.
\par

The rest part of the paper is arranged as follows. The main results
related to semiclassical limit and relaxation time limit are
presented in section 2, the proofs are established in section 3.
 \par\bigskip

\noindent{\bf Notations:\em } $C$ or $c$ always denote the generic
positive constants. $L^2(\R^3)$ is the space of square integral
functions on $\R^3$ with the norm $\parallel\cdot\parallel$ or
$\parallel\cdot\parallel_{L^2(\R^3)}$. $H^k(\R^3)$ with integer
$k\geq 1$ denotes the usual Sobolev space of function $f$ satisfying
$\partial_x^i f\in L^2(\R^3)(0\leq i\leq k)$ with norm
$$\parallel f\parallel_k=\sqrt{\sum\limits _{\small{0\leq
|\alpha|\leq k}}\parallel D^\alpha f\parallel^2},$$ here and after
$\alpha\in N^3,D^\alpha =\partial_{x_1}^{s_1}\partial_{x_2}^{s
_2}\partial_{x_3}^{s _3}$ for $|\alpha |=s_1+s _2+s_3$, Especially
$\parallel\cdot\parallel_0=\parallel\cdot\parallel$. Let $\cal{B}$
be a Banach space, $C^k([0,t];\cal{B})$ denotes the space of
$\cal{B}$-valued k-times continuously differentiable functions on
[0,t]. We can extend the above norm to the vector-valued function
$u=(u_1,u_2,,u_3)$ with $|D^\alpha u|^2=\sum
\limits_{r=1}^3{|D^\alpha u_r|^2}$ and
$$
\parallel D^ku\parallel^2
=\int_{\R^3}(\sum\limits_{r=1}^3\sum\limits_{|\alpha|=k}(D^\alpha
u_r)^2)dx,
$$
and $\parallel u\parallel_k=\parallel u\parallel_{H^k(\R^3)}=\sum
\limits_{i=0}^k\parallel D^iu\parallel$, $\parallel
f\parallel_{L^\infty ([0,T];\cal{B})}=\sup\limits_{0\leq t\leq
T}\parallel f(t)\parallel_{\cal{B}}$. We also use the space
$\mathcal{H}^{k}(\R^3)=\{f\in L^6(\R^3),Df\in H^{k-1}(\R^3)\},k\geq
1$. Sometimes we use $\parallel (.,.,...)\parallel_{H^k(\R^3)}$ or
 $\parallel(.,.,...)\parallel_k$ to denote the norm of the space
$H^k(R^3)\times H^k(\R^3)\times\cdot\cdot\cdot \times H^k(\R^3)$ and
the $\mathcal{H}^{k}(\R^3)$ as well.


\section{Main results  and Preliminary}
\setcounter{equation}{0}

\subsection{Main results}
We consider the initial value problem for the quantum
system~\ef{1.1}--\ef{1.3} with following initial data
\be
  (\rho_i ,u_i)(x,0)=(\rho_{i_0},u_{i_0})(x),~~
 \rho_{i_0}(x)\rightarrow \rho_i^\ast,~u_{i_0}(x)\rightarrow0,
 ~|x|\rightarrow +\infty    \label{1.4}
\ee
with $i=a,b$.  From now on, we set the scaled Debye length to be one
$\lambda =1$ for simplicity.

First of all, we have the global existence and uniqueness theory of
the IVP problem for the quantum system~\ef{1.1}--\ef{1.3} and
\ef{1.4}.

\begin{theorem}\label{2.6}{\rm({\bf Global existence})}
Let the parameters $\eps>0,\tau>0$ be fixed. Assume $P_a$, $P_b\in
C^5(0,+\infty)$ and $\mathcal{C}(x)= c^*$ is a constant satisfying
for two positive constants $\rho_a^\ast,\rho_b^\ast$ that
\be
\rho_a^\ast-\rho_b^\ast-c^*=0, ~~P_a^\prime(\rho_a^\ast),
~~P_b^\prime(\rho_b^\ast)>0,\label{2.1}
\ee
Suppose $\rho_{a_0}>0,\rho_{b_0}>0$ and
 $\sqrt{\rho_{a_0}}-\sqrt{\rho_a^\ast},
  \sqrt{\rho_{b_0}}-\sqrt{\rho_b^\ast},
  u_{a_0},u_{b_0})\in {H^6(\R^3)}\times{\mathcal{H}^5(\R^3)}$.
Then, there is $\Lambda_1>0$ so that if $
\Lambda_0:=\|(\sqrt{\rho_{a_0}}-\sqrt{\rho_a^\ast},
\sqrt{\rho_{b_0}}-\sqrt{\rho_b^\ast},u_{a_0},u_{b_0})
 \|_{H^6\times\mathcal{H}^5(\R^3)}\leq\Lambda_1,$
the unique solution
$(\rho_a^\eps,\rho_b^\eps,u_a^\eps,u_b^\eps,E^\eps)$ of the
IVP~problem $(\ref{1.1})-(\ref{1.3})$ and \qef{1.4} exists globally
in time  with $\rho_a^\eps,\rho_b^\eps>0$ and satisfies
$$
 ( \rho_i^\eps - \rho_a^\ast, \,E^\eps )\in C^k(0,T;H^{6-2k}(\R^3)), \
 u_i^\eps\in C^k(0,T;\mathcal{H}^{5-2k}(\R^3)), \
 k=0,1,2
$$
and
$$
  \|(\rho_a^\eps -\rho_a^\ast,\rho_b^\eps -\rho_a^\ast)\|_{L^\infty(\R^3)}
 +\|E^\eps\|_{L^\infty(\R^3)}
 +\|(u_a^\eps,u_b)\|_{L^\infty(\R^3)}\longrightarrow 0,
$$
as time tends to infinity.
\end{theorem}
\begin{remark} Unlike the unipolar quantum hydrodynamical model
\cite{JL2003,LM215-247,HLM2003,HLMO2003}, we can not get the
exponential convergence to the asymptotical equilibrium state for
bipolar quantum model due to the coupling and cancelation of two
carriers. Usually, the optimal decay rate is algebraic and is left
for the further research \cite{ZLZ2006}
\end{remark}

We then state  semiclassical limit $\eps\to0_+$ of the global in
time solutions to the IVP~\ef{1.1}--\ef{1.3} and \ef{1.4} for any
fixed momentum relaxation time $\tau>0$.
\begin{theorem}\label{semi-limit}{\rm({\bf Global semiclassical limit})}
Let $\tau=1$ and
$(\rho_a^\eps,\rho_b^\eps,u_a^\eps,u_b^\eps,E^\eps)$ be the solution
of the IVP~problem \qef{1.1}--\qef{1.3} and \qef{1.4} given by
Theorem~\ref{2.6}. Then, there is $(\rho_a,u_a,\rho_b,u_b,E)$ with
$\rho_a>0,\rho_b>0$ so that as the Planck constant
$\varepsilon\rightarrow 0$, it holds
$$
 \rho_i^\eps \rightarrow \rho_i ~~\mbox{strongly in}~~
C(0,T;C_b^3\cap H_{loc}^{5-s}); ~~u_i^\eps\rightarrow
u_i~~\mbox{strongly in}~~ C(0,T;C_b^3\cap
\mathcal{H}_{loc}^{5-s});$$
$$E^\eps\rightarrow E~~\mbox{strongly in}~~
C(0,T;C_b^4\cap \mathcal{H}_{loc}^{6-s}),\quad s\in
(0,\mbox{$\frac{1}{2}$}).
$$
for any $T>0$, i=a,b. Note here that  $(\rho_i,u_i,E)$ with $i=a,b$
is the global in-time solution of IVP problem of the bipolar
hydrodynamic model \qef{1.5}-\qef{1.7} and \qef{1.4}.
\end{theorem}

Finally, we consider the combined semiclassical and relaxation
limits for the quantum hydrodynamical model~\ef{1.1}-\ef{1.3}. To
this end, we consider indeed the initial value problem for the
re-scaled system~\ef{1.8}--\ef{1.10} together with the following
initial data
 \be
(\rho_i^\tau,u_i^\tau)(x,0):=(\rho_{i_0}^\tau,u_{i_0}^\tau)
=(\rho_{i_0},\frac{u_{i_0}}{\tau})(x).\label{1.11'}
 \ee
It is easy to verify that there is a unique global in-time strong
solution $(\rho_i^{(\tau,\varepsilon)},u_i^{(\tau,\varepsilon)},
E^{(\tau,\varepsilon)})$ with $i=a,b$  for the IVP problem
\ef{1.8}--\ef{1.10} and \ef{1.11'} based on the Theorem~\ref{2.6}
and the diffusion scaling \ef{scaling}. What left is to establish
the uniform estimates with respect to the parameters
$\eps>0,\tau>0$ in order to pass into the limits. We have,
\begin{theorem}{\rm({\bf Global relaxation and semiclassical limits})}
\label{combine-limit} Let
$(\rho_i^{(\tau,\varepsilon)},u_i^{(\tau,\varepsilon)},
E^{(\tau,\varepsilon)})$ with $i=a,b$ be the unique global solution
of the bipolar QHD equations \qef{1.8}-\qef{1.10} and \qef{1.11'}
obtained in Theorem~\ref{2.6}, then there exist $(\rho_a,\rho_b,E)$
such that as $\varepsilon\rightarrow0$ and $\tau\rightarrow0$
$$
\rho_i^{(\tau,\varepsilon)}\rightarrow\rho_i~~\mbox{strongly in}~~
C(0,T;C_b^2\cap H_{loc}^{4-s}(R^3)),$$
$$E^{(\tau,\varepsilon)}\rightarrow E~~\mbox{strongly
in}~~ C(0,T;C_b^3\cap\mathcal{ H}_{loc}^{5-s}(R^3)),$$
$$
\tau^2|u_i^{(\tau,\varepsilon)}|^2\rightarrow0~~ \mbox{strongly
in}~~L^1(0,T;W_{loc}^{3,3}(R^3)),~~s\in(0,\mbox{$\frac{1}{2}$}).
$$
and $(\rho_a,\rho_b,E)$ is the strong solution of the IVP problem of
bipolar Drift-Diffusion system $(\ref{1.13})$-$(\ref{1.14})$ and
initial data $(\rho_a,\rho_b)(x,0)=(\rho_{a0},\rho_{b0})$.
\end{theorem}
\begin{remark} Although we only state the combined relaxation and
semiclassical limits for the quantum hydrodynamical model
\qef{1.8}--\qef{1.10} here, we claim that the analysis made here
does not require any (communication) restriction between
$\varepsilon$ and $\tau$. That is, one can fix any of the two
parameters $\varepsilon$ and $\tau$ and let the other tend to zero.
Moreover, our analysis for the bipolar model \qef{1.8}--\qef{1.10}
can be applied to justify the semiclassical limit and relaxation
limit for the unipolar model~\cite{JL2005,HLM2003,HLMO2003}.
\end{remark}
\begin{remark}
Although we have only taken the steady state of constant solution in
the profile in above theorems, we claim that our analysis below is
valid for general subsonic steady state.\end{remark}

\subsection{Some lemmas}

\begin{lemma}\label{lem2.1}
Let $f\in H^s(R^3),s\geq\frac{3}{2}$. There is a unique solution of
the divergence equation \be\nabla\cdot u=f,~~~\nabla\times
u=0,~~~u(x)\rightarrow 0,~~~|x| \rightarrow+\infty.\nnm\ee
satisfying \be\parallel u
\parallel_{L^6(R^3)}\leq C\parallel f \parallel_{L^2(R^3)},
~~~\parallel Du\parallel_{H^s(R^3)}\leq C\parallel
f\parallel_{H^s(R^3)}.\nnm\ee
\end{lemma}
\begin{lemma}\label{lem2.2}
Let $f\in H^s(R^3),s\geq\frac{3}{2}$ with $\nabla\cdot f=0$. There
is a unique solution $u$ of the vorticity equation
$$\nabla\times u=f,~~~\nabla\cdot u=0,~~~u(x)\rightarrow 0,~~~|x|\rightarrow
+\infty.$$
 satisfying
 $$\parallel u \parallel_{L^6(R^3)}\leq C\parallel f \parallel_{L^2(R^3)},
~~~\parallel Du\parallel_{H^s(R^3)}\leq C\parallel
f\parallel_{H^s(R^3)}. $$
\end{lemma}

We will also use the Moser type calculus lemmas.
\begin{lemma}\label{lem2.3}
Let $f,g \in H^s(R^3)\bigcap L^\infty(R^3)$, then it holds
 $$
 \parallel D^\alpha (fg)\parallel
 \leq
 C\parallel g\parallel_{L^\infty}\cdot\parallel D^\alpha
 f\parallel
 +C\parallel f\parallel_{L^\infty}
 \cdot\parallel D^\alpha g\parallel$$
 for $\alpha\in N^3,1\leq |\alpha|\leq s ,s\geq 0 $ is an
 integer.
 \end{lemma}
 \begin{lemma} \label{lem2.4}
Let $f\in H^s(R^3)$ with $s\geq 0$ be an integer  and function
$F(\rho)$ smooth enough and $F(0)=0$ then $F(f)(x)\in H^s(R^3)$ and
$$\parallel F(f)\parallel_{H^s(R^3)}\leq C\parallel
f\parallel_{H^s(R^3)}.~$$

\end{lemma}

\section{The proof of  main results}
\setcounter{equation}{0}

 The local in-time existence result of QHD
model has been obtained in \cite{LM215-247,HLMO2003}. The framework
used there is to study an extended problem derived based on a
deposition of the original problem, which in turn implies the
expected problem as a special case. The method employed in
\cite{LM215-247,HLMO2003} can be applied to our bipolar model
directly. The proof is straightforward, and we have
\begin{lemma}{\rm({\bf Local existence})} \label{local}
Let the parameters $\eps>0,\tau>0,\lambda>0$ be fixed. Assume that
there are constants $\rho_a^\ast,\rho_b^\ast>0$ and $c*$ satisfying
$\rho_a^\ast-\rho_b^\ast-c*=0$, $\mathcal{C}(x)-c*\in H^5(R^3)$, and
$P_a$, $P_b\in C^5(0,+\infty)$. Assume
$(\sqrt{\rho_{i_0}}-\sqrt{\rho_i^\ast},u_{i_0})\in
H^6(R^3)\times\mathcal{H}^5(R^3)$ with $\rho_{i_0}>0$, then there
exists a finite time $T^*>0$ such that the unique solution
$(\rho_a,\rho_b,u_a,u_b,E)$ with $\rho_{a}>0,\rho_{b}>0$ of the
problem $(\ref{1.1})$-$(\ref{1.3})$ and $(\ref{1.11'})$ exists in
$[0,T^*]$, and it satisfies
$$\rho_i-\rho_i^\ast\in C^k([0,T^*];H^{6-2k}(R^3)),
~u_i\in C^k([0,T^*];\mathcal{H}^{5-2k}(R^3)),~k=0,1,2.$$
$$E\in C^k([0,T^*];\mathcal{H}^{6-k}(R^3)),~k=0,1.$$
\end{lemma}

Here, we also mention the global existence theory for the quantum
hydrodynamical model and bipolar hydrodynamical model. The
well-posedness of steady state subsonic solutions has been proved
also in \cite{{J463-479},{JL2004},{zJ845-856}}. Transient solutions
are shown to exist either locally in time
\cite{{HJ2003,1-15},{HLM2003}} or globally in time for data close to
a steady state \cite{{HLMO2003},{JL2003},{JL2004}, {LM215-247}}. The
bipolar hydrodynamic(HD) model of the global solutions has been
studied in \cite{GHL2003}.
\par

\subsection{Reformulation of original problem}

In this section, we study the global solutions and the asymptotic
limits with the case $\mathcal{C}(x)=c^*$. Inspired by \cite{JL2005}
we consider the problem when the initial data of
$(\rho_i^\tau,u_i^\tau,E^\tau)$ is around the steady state
$(\rho_i^\ast,0,0)$ and make use of energy estimates to analyze
perturbation of the global in-time solutions. To this end, we employ
the fourth-order wave equations for $\sqrt{\rho_i^\tau}$ and the
equation of the vorticity of velocity $u_i^\tau$. The poisson
equation is used to deal with the coupling of the two carriers and
some technique is used to deal with the smallness both of
$\varepsilon$ and $\tau$.
\par

Since we are interested in not only the global existence theory but
also the asymptotical analysis of strong solutions with respect to
small parameters, we deal with the scaled IVP problem
\ef{1.8}--\ef{1.10} and \ef{1.11'} directly. Because the scaled
scaled IVP problem \ef{1.8}--\ef{1.10} and \ef{1.11'} is equivalent
to the original IVP problem \ef{1.1}--\ef{1.3} and \ef{1.4} for
strong (classical) solutions.  For simplicity, we take $\lambda=1$
and let $(.)_t$ denote $\partial_t(.)$ and omit the index
$\eps,\tau$ to simplify the presentation in the following argument.
From (\ref{1.8})-(\ref{1.10}) and \ef{1.11'} the equations for
$\psi_i=\sqrt{\rho_i^\tau} $ with $u_i=u_i^\tau(i=a,b)$ can be
obtained  \bma
 \tau^2\psi_{itt}+\psi_{it}+&\frac{\varepsilon^2\triangle^2\psi_i}{4}
 +\frac{q_i}{2\psi_i}\nabla\cdot(\psi_i^2E)
-\frac{1}{2\psi_i}\nabla^2(\psi_i^2 u_i\otimes u_i)
\nnm \\
-&\frac{1}{2\psi_i}\triangle
P_i(\psi_i^2)+\frac{\psi_{it}^2}{\psi_i}
-\frac{\varepsilon^2|\triangle\psi_i|^2}{4\psi_i}=0,
\label{(3.1)}
\ema
with the initial value
$$
\psi_i(x,0):=\psi_{i_0}(x)=\psi_{i_0}^\tau(x)=\sqrt{\rho_{i_0}(x)},$$
$$
\psi_{it}(x,0):=\psi_{i1}(x)=-\frac{1}{2}\psi_{i_0}^\tau\nabla\cdot
u_{i_0}^\tau-u_{i_0}^\tau\cdot
  \nabla\psi_{i_0}^\tau.
$$
Also from (\ref{1.8})-(\ref{1.11'}) with the fact
$(u_i\cdot\nabla)u_i=\frac{1}{2}\nabla(|u_i|^2)-u_i\times(\nabla
\times u_i),$ the equations for $u_i=u_i^\tau$ (i=a,b)
 \be
\tau^2u_{it}+u_i+\frac{\tau^2}{2}\nabla({|u_i|}^2)-
\tau^2u_i\times\phi_i+\frac{\nabla (\psi_i^2)}{\psi_i^2}=
q_iE+\frac{\varepsilon
^2}{2}\nabla(\frac{\triangle\psi_i}{\psi_i}), \label{3.2}
 \ee
where $\phi_i=\nabla\times u_i$ denotes the vorticity of $u_i$.
Taking curl of (3.2), we have
\be
\tau^2\phi_{it}+\phi_i+\tau^2(u_i\cdot\nabla)\phi_i
+\tau^2\phi_i\nabla\cdot
u_i-\tau^2(\phi_i\cdot\nabla)u_i=0,\label{3.3}
 \ee
Introduce new variables $w_i=\psi_i-\sqrt{\rho_i^*}$ with $i=a,b$,
then the system for $(w_a,w_b,\phi_a,\phi_b,E)$ is
\bgr
\tau^2
 w_{att}+w_{at}+\frac{\varepsilon^2\triangle^2w_a}{4}
 +\frac{1}{2}(w_a+\sqrt{\rho_a^*})
 \nabla\cdot E-P_a^\prime (\rho_a^*)\triangle w_a=f_{a1},\label{3.4}
\\
 \tau^2 w_{btt}+w_{bt}+\frac{\varepsilon^2\triangle^2w_b}{4}
 -\frac{1}{2}(w_b+\sqrt{\rho_b^*})\nabla\cdot E
 -P_b^\prime (\rho_b^*)\triangle w_b=f_{b1},\label{3.5}
\\
 \tau^2\phi_{at}+\phi_a=f_{a2},\label{3.6}\\
 \tau^2\phi_{bt}+\phi_b=f_{b2},\label{3.7}
\egr
and
 \be
   \nabla\cdot E
 = w_a^2-w_b^2+2\sqrt{\rho_a^*}w_a
   -2\sqrt{\rho_b^*}w_b,~~\nabla\times E=0,\label{3.8}
\ee
 where
 \bma
 f_{i1}:=f_{i1}(x,t)
 = &\frac{-\tau^2w_{it}^2}{w_i+\sqrt{\rho_i^*}}-q_i\nabla w_iE
  +(P_i^\prime((w_i+\sqrt{\rho_i^*})^2)
  -P_i^\prime(\rho_i^*))\triangle w_i\nnm
  \\
  &+2(w_i+\sqrt{\rho_i^*})
    P_i^{\prime\prime}((w_i+\sqrt{\rho_i^*})^2)|\nabla w_i|^2\nnm
  \\
  &+\frac{\varepsilon^2(\triangle w_i)^2}{4(w_i+\sqrt{\rho_i^*})}
   +\frac{\tau^2\nabla^2((w_i+\sqrt{\rho_i^*})^2u_i\otimes
   u_i)}{2(w_i+\sqrt{\rho_i^*})},\label{3.9}\\
 f_{i2}:=f_{i2}(x,t)
 =&\tau^2((\phi_i\cdot\nabla)u_i-(u_i\cdot\nabla)\phi_i
   -\phi_i\nabla\cdot u_i),\quad i=a,b.\label{3.10}
 \ema
The last term in (\ref{3.9}) can be decomposed by using equation
(\ref{1.8}) as
\bma
&\frac{\tau^2\nabla^2((w_i+\sqrt{\rho_i^*})^2u_i\otimes
   u_i)}{2(w_i+\sqrt{\rho_i^*})}
   \nnm\\
=&\tau^2\{-w_{it}\nabla\cdot u_i-2u_i\cdot\nabla w_{it}
-\frac{w_{it}u_i\cdot\nabla w_i}{2(w_i+\sqrt{\rho_i^*})}+\nabla
w_i\cdot((u_i\cdot\nabla)u_i)\nnm\\
&+\frac{(w_i+\sqrt{\rho_i^*})}{2}\sum_{k,l=1}^3|\partial_ku_i^l|^2
-\frac{(w_i+\sqrt{\rho_i^*})}{2}|\phi_i|^2-u_i\cdot\nabla(u_i\cdot\nabla
w_i)\nnm\\
&+\frac{1}{2(w_i+\sqrt{\rho_i^*})}(w_{it}+u_i\cdot\nabla
w_i)(u_i\cdot\nabla w_i)\},\quad i=a,b.\label{3.11} \ema
 The initial
conditions for (\ref{3.4})-(\ref{3.7}) are \bma
&w_i(x,0):=w_{i_0}(x)=\psi_{i_0}-\sqrt{\rho_i^*},~~
\phi_i(x,0):=\phi_{i_0}(x)=\frac{1}{\tau}\nabla \times
u_{i_0}(x),\nnm\\
&w_{it}(x,0):=w_{i1}(x)=\frac{1}{\tau}(-u_{i_0}\cdot\nabla w_{i_0}
-\frac{1}{2}(w_{i_0}+\sqrt{\rho_i^*})\nabla\cdot u_{i_0}),\quad
i=a,b.\nnm
\ema
We will also use the relation between $\nabla\cdot u_i$ and $\nabla
w_i,w_{it}$
 \be
 2w_{it}+2u_i\cdot\nabla
w_{i}+(w_{i}+\sqrt{\rho_i^*})\nabla\cdot u_i=0,\quad
i=a,b.\label{3.12}
\ee

\subsection{The a-priori estimates  }
In this section, we will mainly study the reformulated equations
 (\ref{3.4})-(\ref{3.8}) in order to obtain the a-priori estimates of
$w_a,w_b,\phi_a,\phi_b,E.$
\par Set the workspace as
$$
\textsl{X}(T)=\{(w_a,w_b,u_a,u_b)
\in
L^\infty([0,T];(H^6(R^3))^2\times(\mathcal{H}^5(R^3))^2\}
$$
and assume the quantity
 \bma
 \delta_T=&\max_{0\leq t\leq T}\{\parallel(w_a,w_b)(.,t)\parallel_4^2
   +\parallel\tau(\partial_tw_a,\partial_tw_b)(.,t)
   \parallel_3^2+\parallel\tau( u_a, u_b)(.,t)
   \parallel_{\mathcal{H}^{4}}^2\}\nnm\\
&+\int_0^T\{\parallel(u_a,u_b)(.,t)\parallel_{\mathcal{H}^{3}}^2+
   \parallel(w_a,w_b)(.,t)\parallel_5^2
   +\parallel E(.,t)\parallel_{\mathcal{H}^{2}}^2\}dt,\label{3.13}
   \ema
is small, then by Sobolev embedding theorem we know that the
sufficiently small $\delta_T$ can assure the positivity of
$\psi_a,\psi_b$ as
$$\frac{\sqrt{\rho_a^*}}{2}\leq
w_a+\sqrt{\rho_a^*}\leq\frac{3}{2}\sqrt{\rho_a^*},~~
\frac{\sqrt{\rho_b^*}}{2}\leq
w_b+\sqrt{\rho_b^*}\leq\frac{3}{2}\sqrt{\rho_b^*}$$ By Sobolev
embedding theorem, from the assumption for $\delta_T$, we also have
\bgr
\parallel(D^\alpha w_a,D^\alpha w_b,\tau D^\beta
w_{at},\tau D^\beta w_{bt})\parallel_{L^\infty (R^3\times[0,T])}\leq
c\delta_T ,~~|\alpha|\leq2,|\beta|\leq1.\label{3.14}
\\
\parallel(\tau D^\alpha u_a,\tau D^\alpha
u_b)\parallel_{L^\infty (R^3\times[0,T])}\leq c\delta_T
,~~~|\alpha|\leq2.\label{3.15}\\ \int_0^T\parallel(D^\alpha
u_a,D^\alpha u_b,\tau^2 u_{at}, \tau^2
u_{bt})(.,t)\parallel_{L^\infty (R^3)}^2dt\leq
c\delta_T,~~~|\alpha|\leq1.\label{3.16}
 \egr
The last inequality (\ref{3.16}) is obtained from the equations for
$u_a,u_b$ and Sobolev embedding theorem, the assumption for
$\delta_T$. The $c$ or $C$ denote the generic positive constant and
does not necessarily be the same here and after. Using
Lemma~\ref{lem2.1}, from the poisson equation (\ref{3.8}) we have
\be
\parallel E\parallel_{L^\infty([0,T];\mathcal{H}^{5}(R^3))}\leq
c\delta_T, ~~\parallel D^\alpha
E\parallel_{L^\infty(R^3\times[0,T])}\leq
c\delta_T,~~|\alpha|\leq3.\label{3.17} \ee
 Next, we will establish energy estimates to extend the
solution to global one.
\par

We have the main a-priori estimate lemma.
\begin{lemma} \label{lem3.2} Suppose $(w_a,w_b,u_a,u_b,E)(x,t)$
is local solution with $\delta_T \ll1$, then it holds
 \be
E_1(t)+\int_0^tE_2(s)ds\leq c\Lambda_0,\label{3.18}\ee
 for
$t\in(0,T)$ and $c>0$ is a constant independent of $\varepsilon $
and $\tau$. The $\Lambda_0$ is defined in Theorem 2.6, and here
\bma
 E_1(t):=\{\parallel(&w_a,w_b)(.,t)\parallel_4^2
+(\tau+\varepsilon^2)\parallel(D^5w_a,D^5w_b)(.,t)\parallel^2\nnm\\
+&\tau\varepsilon^2\parallel(D^6w_a,D^6w_b)(.,t)\parallel^2
+\tau^2\parallel(w_{at},w_{bt})(.,t)\parallel_3^2\nnm\\
+&\tau^3\parallel(D^4w_{at},D^4w_{bt})(.,t)\parallel^2
+\tau^2\parallel(u_a,u_b)(.,t)\parallel_{\mathcal{H}^4}^2\nnm\\
+&\tau^3\parallel(D^5u_a,D^5u_b)(.,t)\parallel^2+\parallel
E(.,t)\parallel_{\mathcal{H}^5}^2\},\nnm\\[2mm]
E_2(t):=\{\parallel(&\nabla w_a,\nabla
w_b)(.,t)\parallel_4^2
+\varepsilon^2\parallel(D^6w_a,D^6w_b)(.,t)\parallel^2\nnm\\
+&\parallel(w_{at},w_{bt})(.,t)\parallel_3^2
+\tau\parallel(D^4w_{at},D^4w_{bt})(.,t)\parallel^2
\nnm\\
+&\parallel(u_a,u_b)(.,t)\parallel_{\mathcal{H}^4}^2
+\tau\parallel(D^5u_a,D^5u_b)(.,t)\parallel^2 +\parallel
E(.,t)\parallel_{\mathcal{H}^5}^2\}.\nnm \ema
\end{lemma}

\textbf{Proof}:\ \
 \textbf{Step 1. The estimates for $w_a,w_b$.}\\
\textit{Step 1.1. basic estimates}. Assume $\tau<1$ for simplicity.
Multiplying (\ref{3.4}) by $(w_a+2w_{at})$ and (\ref{3.5}) by
$(w_b+2w_{bt}),$ integrating by parts the resulted equations over
$R^3$, summing the resulted two equalities and noticing the facts
from equation (\ref{3.8})
 \bma
&\int_{R^3}\{(\frac{1}{2}(w_a+\sqrt{\rho_a^*})\nabla\cdot E)w_a
   -(\frac{1}{2}(w_b+\sqrt{\rho_b^*})\nabla\cdot E)w_b
   \}dx\nnm\\
&=\ \frac{1}{4}\int_{R^3}|\nabla\cdot
   E|^2dx-\frac{1}{4}\int_{R^3}\nabla(w_a^2-w_b^2)\cdot E
   dx,\nnm
\ema
 and
\bma &\int_{R^3}\{\frac{1}{2}((w_a+\sqrt{\rho_a^*})\nabla\cdot E)
   2w_{at}-\frac{1}{2}((w_b+\sqrt{\rho_b^*})\nabla\cdot E)
   2w_{bt}\} dx \nnm\\
&=\frac{1}{4}\frac{d}{dt}\int_{R^3}|\nabla\cdot E|^2dx,\nnm
\ema
then after a tedious but straightforward calculation we have
\bma
 \frac{d}{dt} \int_{R^3}\{\tau ^2w_{at}^2+ \tau ^2&w_aw_{at}+\frac{w_a^2}{2}
+\tau ^2w_{bt}^2+ \tau ^2w_bw_{bt}+\frac{w_b^2}{2}+
P_a^\prime(\rho_a^*)|\nabla w_a|^2\nnm\\
   +P_b^\prime(\rho_b^*)&|\nabla
w_b|^2+\frac{\varepsilon^2}{4}(|\triangle w_a|^2
   +|\triangle w_b|^2)+\frac{1}{4}|\nabla\cdot E|^2\}dx
   \nnm\\
   + \int_{R^3}\{(2-\tau ^2)(w_{at}^2&+w_{bt}^2)
   + P_a^\prime(\rho_a^*)|\nabla w_a|^2
   + P_b^\prime(\rho_b^*)|\nabla w_b|^2
   +\frac{\varepsilon^2}{4}(|\triangle w_a|^2
   +|\triangle w_b|^2)\nnm\\
   +\frac{1}{4}|\nabla\cdot E|^2\}&dx\nnm\\
  =\frac{1}{2}\int_{R^3}(w_a\nabla w_a-
   &w_b\nabla w_b)\cdot E dx\nnm\\
   +\int_{R^3}\{f_{a1}(x,t)(&w_a+2w_{at})
   +f_{b1}(x,t)(w_a+2w_{bt})\}dx ,\label{3.19}
   \ema
The right-hand side of (\ref{3.19}) can be analyzed as follows. By
Sobolev embedding theorem and H\"{o}lder inequality, Young's
inequality \bma \int_{R^3}w_i\nabla w_i\cdot E dx\leq&
\parallel w_i\parallel_{L^3}\parallel \nabla
w_i\parallel_{L^2}\cdot\parallel  E
\parallel_{L^6}
\nnm\\
\leq &c(\parallel w_i\parallel_{L^2}+\parallel \nabla
w_i\parallel_{L^2}) (\parallel \nabla
w_i\parallel_{L^2}\cdot\parallel  E
\parallel_{L^6})\nnm\\
\leq &c(\delta_T)^{\frac{1}{2}}(\parallel \nabla
w_i\parallel^2+\parallel \nabla\cdot E
\parallel^2),\label{3.20}\ema
i=a,b. Here we have used Lemma~\ref{lem2.1} to estimate $\parallel
DE\parallel^2$ by $\parallel \nabla\cdot E\parallel^2$.
  Some other key terms of the right-hand side are analyzed as
\be
\int_{R^3}[P_i^\prime((w_i+\sqrt{\rho_i^*})^2)-P_i^\prime(\rho_i^*)]
\triangle w_i\cdot(2w_{it})dx\leq
c(\delta_T)^{\frac{1}{2}}(\parallel\triangle
w_i\parallel^2+\parallel w_{it}\parallel^2),\label{3.21} \ee
\be\int_{R^3}\tau^2u_i\cdot\nabla w_{it}(2w_{it})dx
=-\int_{R^3}\tau^2\nabla\cdot u_i(w_{it})^2dx\leq
c(\delta_T)^{\frac{1}{2}}\parallel w_{it}\parallel^2,\label{3.22}
\ee \be\int_{R^3}\tau^2u_i\nabla(u_i\cdot\nabla w_i)(2w_{it})dx
\leq c(\delta_T)^{\frac{1}{2}}\parallel(\nabla w_i,\triangle
w_i,w_{it},\phi_i)\parallel^2,\label{3.23} \ee In (\ref{3.23}) we
have used $\parallel Du_i\parallel^2\leq c(\parallel\nabla\cdot
u_i\parallel^2+\parallel\nabla\times u_i\parallel^2),$ and
$\parallel\nabla w_i\parallel^2,\parallel w_{it}\parallel^2$ to
estimate $\nabla\cdot u_i$ through equation (\ref{3.12}). Then, by
(\ref{3.19})-(\ref{3.23}), using integration by parts, H\"{o}lder
inequality, Young's inequality and the Moser type
Lemma~\ref{lem2.3}, Lemma \ref{lem2.4} to estimate the other terms
of the right-hand side of (\ref{3.19}), we can arrive at
 \bma \frac{d}{dt} \int_{R^3}\{&\tau ^2w_{at}^2+
\tau ^2w_aw_{at}+\frac{w_a^2}{2} +\tau ^2w_{bt}^2+ \tau
^2w_bw_{bt}+\frac{w_b^2}{2}+ P_a^\prime(\rho_a^*)|\nabla
w_a|^2\nnm\\
&+P_b^\prime(\rho_b^*)|\nabla
w_b|^2+\frac{\varepsilon^2}{4}(|\triangle w_a|^2
   +|\triangle w_b|^2)+\frac{1}{4}|\nabla\cdot E|^2\}dx
    \nnm\\
+ \int_{R^3}\{(&2-\tau ^2)(w_{at}^2+w_{bt}^2)
   + P_a^\prime(\rho_a^*)|\nabla w_a|^2
   + P_b^\prime(\rho_b^*)|\nabla w_b|^2
   +\frac{\varepsilon^2}{4}(|\triangle w_a|^2
   +|\triangle w_b|^2)~~
    \nnm\\
    &+\frac{1}{4}|\nabla\cdot E|^2\}dx\nnm\\
 \leq
 ~c(\delta_T&)^{\frac{1}{2}}\parallel(\nabla w_a,\nabla w_b,w_{at},w_{at},
 \nabla\cdot E,\phi_a,\phi_b)\parallel^2+c(\delta_T)^{\frac{1}{2}}
 \parallel(\triangle w_a,\triangle
 w_b)\parallel^2,\label{3.24}
 \ema
The right hand side of estimate (\ref{3.24}) will be used later in
the closure of the a-priori estimates.
\par

\textit{Step 1.2.  the higher-order estimates for $w_a,w_b$.}
 Differentiate (\ref{3.4}) and
(\ref{3.5}) with respect to $x$, then the functions
$\widetilde{w_a}:=D^\alpha w_a,\widetilde{w_b}:=D^\alpha w_b $ and
$\widetilde{E}:=D^\alpha E$$(1<|\alpha|\leq3)$\footnote[1]{ we can
first assume the solution $(w_a,w_b,u_a,u_b)$ has higher order
regularity so that we can take derivatives since the final
a-priori estimation  will be still valid for these solutions by
applying the Friedrich mollifier
 to $(w_a,w_b,u_a,u_b)$ .} satisfy
\bma
&\tau^2\widetilde{w_i}_{tt}+\widetilde{w_i}_{t}+\frac{\varepsilon^2}{4}
\triangle^2\widetilde{w_i} +
\frac{q_i}{2}(w_i+\sqrt{\rho_i^*})\nabla\cdot\widetilde{E}
-P_i^\prime(\rho_i^*)\triangle\widetilde{w_i}\nnm\\
=&D^\alpha f_{i1}(x,t)-D^\alpha(\frac{q_i}{2}(w_i+\sqrt{\rho_i^*}
) \nabla\cdot E) +\frac{q_i}{2}(w_i+\sqrt{\rho_i^*}
)\nabla\cdot\widetilde{E}\nnm\\
\stackrel{def}{=}&F_{i}(x,t),(i=a,b,~
q_a=1,q_b=-1.)\label{3.25}\ema
 Multiplying (\ref{3.25}) for $i=a$ by
$(\widetilde{w_a}+2\widetilde{w_a}_t)$, and (\ref{3.25}) for $i=b$
by $(\widetilde{w_b}+2\widetilde{w_b}_t)$, integrating by parts
 over $R^3$, summing the resulted equalities, also noticing the
 facts
\bma &\int_{R^3}\{\frac{1}{2}(w_a+\sqrt{\rho_a^*})
   \nabla\cdot(\widetilde{E})(\widetilde{w_a}+2\widetilde{w_a}_t)
   -\frac{1}{2}(w_b+\sqrt{\rho_b^*})
   \nabla\cdot(\widetilde{E})(\widetilde{w_b}+2\widetilde{w_b}_t)\}dx\nnm\\
 &=\frac{1}{4}\int_{R^3}|\nabla\cdot\widetilde{E}|^2dx
 +\frac{1}{4}\frac{d}{dt}\int_{R^3}|\nabla\cdot\widetilde{E}|^2dx
   - \frac{1}{4}\int_{R^3}\nabla\cdot\widetilde{E}D^\alpha(w_a^2-w_b^2)dx
   \nnm\\
   &\ \ \ -\frac{1}{2}\int_{R^3}\nabla\cdot\widetilde{E}
         D^\alpha(w_a^2-w_b^2)_tdx
   +\int_{R^3}\frac{1}{2}w_a
   \nabla\cdot(\widetilde{E})(\widetilde{w_a}+2\widetilde{w_a}_t)dx,\nnm\\
   &\ \ \ -\int_{R^3}\frac{1}{2}w_b
   \nabla\cdot(\widetilde{E})(\widetilde{w_b}
     +2\widetilde{w_b}_t)dx,\label{3.26}\ema
after a tedious but straightforward computation one can get \bma
&\frac{d}{dt}\int_{R^3}\{\tau^2\widetilde{w_a}_t^2
   +\tau^2\widetilde{w_a}\widetilde{w_a}_t
   +\frac{1}{2}\widetilde{w_a}^2
   +\tau^2\widetilde{w_b}_t^2
   +\tau^2\widetilde{w_b}\widetilde{w_b}_t
   +\frac{1}{2}\widetilde{w_b}^2
   + P_a^\prime(\rho_a^*)|\nabla\widetilde{w_a}|^2\nnm\\
   &~~~~~~~~+P_b^\prime(\rho_b^*)|\nabla\widetilde{w_b}|^2
   +\frac{\varepsilon^2}{4}(|\triangle\widetilde{w_a}|^2
   +|\triangle\widetilde{w_b}|^2)
   +\frac{1}{4}|\nabla\cdot\widetilde{E}|^2\}dx\nnm\\
  &+\int_{R^3}\{(2-\tau^2)(\widetilde{w_a}_t^2+
   \widetilde{w_b}_t^2)+ P_a^\prime(\rho_a^*)|\nabla\widetilde{w_a}|^2+
   P_b^\prime(\rho_b^*)|\nabla\widetilde{w_b}|^2
    \nnm\\
  &~~~~~~~~+\frac{\varepsilon^2}{4}(|\triangle\widetilde{w_a}|^2
   +|\triangle\widetilde{w_b}|^2)+\frac{1}{4}|\nabla\cdot\widetilde{E}|^2 \}dx
   \nnm\\
   =&~~\int_{R^3}\{F_a\cdot(\widetilde{w_a}+2\widetilde{w_{a}}_t)
   +F_b\cdot(\widetilde{w_b}+2\widetilde{w_{b}}_t)\}dx
   +\frac{1}{4}\int_{R^3}\nabla\cdot\widetilde{E}D^\alpha(w_a^2-w_b^2)dx\nnm\\
 &+\frac{1}{2}\int_{R^3}\nabla\cdot\widetilde{E}D^\alpha(w_a^2-w_b^2)_tdx
   -\frac{1}{2}\int_{R^3}w_a\nabla\cdot\widetilde{E}
   (\widetilde{w_a}+2\widetilde{w_{a}}_t)dx\nnm\\
&+\frac{1}{2}\int_{R^3}w_b\nabla\cdot\widetilde{E}
   (\widetilde{w_b}+2\widetilde{w_{b}}_t)dx,
   \label{3.27}\ema
Similar with the analysis of basic estimates, using Moser type
inequality Lemma \ref{lem2.3}, Lemma \ref{lem2.4} and the priori
assumptions (\ref{3.13})-(\ref{3.17}) and using H\"{o}lder
inequality, Young's inequality to estimate the terms of the
right-hand side of (\ref{3.27}), we can arrive at
\bma
&\frac{d}{dt}\int_{R^3}\{\tau^2\widetilde{w_a}_t^2
   +\tau^2\widetilde{w_a}\widetilde{w_a}_t
   +\frac{1}{2}\widetilde{w_a}^2
   +\tau^2\widetilde{w_b}_t^2
   +\tau^2\widetilde{w_b}\widetilde{w_b}_t
   +\frac{1}{2}\widetilde{w_b}^2
   + P_a^\prime(\rho_a^*)|\nabla\widetilde{w_a}|^2\nnm\\
  &~~~~~~~~~~~~~
   +P_b^\prime(\rho_b^*)|\nabla\widetilde{w_b}|^2
   +\frac{\varepsilon^2}{4}(|\triangle\widetilde{w_a}|^2
   +|\triangle\widetilde{w_b}|^2)
   +\frac{1}{4}|\nabla\cdot\widetilde{E}|^2\}dx\nnm\\
  &+\int_{R^3}\{(2-\tau^2)(\widetilde{w_a}_t^2+
   \widetilde{w_b}_t^2)+ P_a^\prime(\rho_a^*)|\nabla\widetilde{w_a}|^2+
   P_b^\prime(\rho_b^*)|\nabla\widetilde{w_b}|^2
    \nnm\\
  &
  ~~~~~~~~~~~~~~~+\frac{\varepsilon^2}{4}(|\triangle\widetilde{w_a}|^2
   +|\triangle\widetilde{w_b}|^2)
   +\frac{1}{4}|\nabla\cdot\widetilde{E}|^2 \}dx
  \nnm\\
 \leq& ~~c\delta_T^{\frac{1}{2}}\parallel(\nabla w_a,\nabla
w_b,w_{at},w_{bt},\phi_a,\phi_b,\nabla \cdot E)\parallel_3^2
 +c\delta_T^{\frac{1}{2}}\parallel(D^5w_a,D^5w_b)\parallel^2,\label{3.28}
 \ema
Note that, we can not deal with the last term in (\ref{3.28}) by the
energy of left-hand side now, so we have to do the highest-order
estimates in different way in order to overcome the difficulty.\par

 \textit{Step 1.3. the highest-order estimates for
$w_a,w_b$.} Taking $|\alpha|=4$, we can get the equations for
$\widetilde{w_a}:=D^\alpha w_a,\widetilde{w_b}:=D^\alpha w_b $ and
$\widetilde{E}:=D^\alpha E$. We also use the form of (\ref{3.25})
for simplicity. This time, using
$(\widetilde{w_a}+2\tau\widetilde{w_a}_t)$ to multiply
$(\ref{3.25})_{i=a}$ and $(\widetilde{w_b}+2\tau\widetilde{w_b}_t)$
to multiply $(\ref{3.25})_{i=b}$ but for $|\alpha|=4$. we can get as
former
\bma &\frac{d}{dt}\int_{R^3}\{\tau^3\widetilde{w_a}_t^2
   +\tau^2\widetilde{w_a}\widetilde{w_a}_t
   +\frac{1}{2}\widetilde{w_a}^2
   +\tau^3\widetilde{w_b}_t^2
   +\tau^2\widetilde{w_b}\widetilde{w_b}_t
   +\frac{1}{2}\widetilde{w_b}^2
   + \tau P_a^\prime(\rho_a^*)|\nabla\widetilde{w_a}|^2\nnm\\
  &~~~~~~~~~~
   +\tau P_b^\prime(\rho_b^*)|\nabla\widetilde{w_b}|^2
   +\frac{\tau \varepsilon^2}{4}(|\triangle\widetilde{w_a}|^2
   +|\triangle\widetilde{w_b}|^2)
   +\frac{\tau }{4}|\nabla\cdot\widetilde{E}|^2\}dx\nnm\\
  &+\int_{R^3}\{(2\tau -\tau^2)(\widetilde{w_a}_t^2+
   \widetilde{w_b}_t^2)+ P_a^\prime(\rho_a^*)|\nabla\widetilde{w_a}|^2+
   P_b^\prime(\rho_b^*)|\nabla\widetilde{w_b}|^2
    \nnm\\
  &~~~
  ~~~~~~~+\frac{\varepsilon^2}{4}(|\triangle\widetilde{w_a}|^2
   +|\triangle\widetilde{w_b}|^2)
   +\frac{1}{4}|\nabla\cdot\widetilde{E}|^2 \}dx
   \nnm\\
   =&\int_{R^3}\{F_a\cdot(\widetilde{w_a}+2\tau \widetilde{w_{a}}_t)
   +F_b\cdot(\widetilde{w_b}+2\tau \widetilde{w_{b}}_t)\}dx
   +\frac{1}{4}\int_{R^3}\nabla\cdot\widetilde{E}D^\alpha(w_a^2-w_b^2)dx\nnm\\
 &+\frac{1}{2}\int_{R^3}\tau
\nabla\cdot\widetilde{E}D^\alpha(w_a^2-w_b^2)_tdx
   -\frac{1}{2}\int_{R^3}w_a\nabla\cdot\widetilde{E}
   (\widetilde{w_a}+2\tau \widetilde{w_{a}}_t)dx
   \nnm\\
&+\frac{1}{2}\int_{R^3}w_b\nabla\cdot\widetilde{E}
   (\widetilde{w_b}+2\tau \widetilde{w_{b}}_t)dx,\label{3.29}
   \ema
In the right-hand side of (\ref{3.29}), the terms multiplied by
$2\tau \widetilde{w_{a}}_t,2\tau \widetilde{w_{b}}_t$ need a
special analysis. Taking $i=a$ for example, the key terms are
analyzed as \bma &\int_{R^3}[P_a^\prime((w_a+\sqrt{\rho_a^*})^2)
-P_a^\prime(\rho_a^*)]\triangle \widetilde{w_a} \cdot2\tau
\widetilde{w_{a}}_tdx\nnm\\
&\leq-\frac{d}{dt}\int_{R^3}\tau[P_a^\prime((w_a+\sqrt{\rho_a^*})^2)
-P_a^\prime(\rho_a^*)]|\nabla\widetilde{w_a}|^2dx+c\delta_T^{\frac{1}{2}}
\parallel\nabla\widetilde{w_a}\parallel^2\nnm\\
&\ \ \ +c\delta_T^{\frac{1}{2}}\tau
\parallel\widetilde{w_{a}}_t\parallel^2,\label{3.30}\ema
\be
\int_{R^3}\tau^2u_a\nabla\widetilde{w_{a}}_t\cdot2\tau\widetilde{w_{a}}_tdx
=-\int_{R^3}\tau^3\nabla\cdot u_a|\widetilde{w_{a}}_t|^2dx\leq
c\delta_T^{\frac{1}{2}}\tau
\parallel\widetilde{w_{a}}_t\parallel^2,\label{3.31}\ee
and \bma&\int_{R^3}\tau^2u_a\nabla(u_a\cdot\nabla\widetilde{w_a}
)\cdot2\tau\widetilde{w_{a}}_tdx\nnm\\
&\leq-\frac{d}{dt}\int_{R^3}\tau(\tau
u_a\cdot\nabla\widetilde{w_a})^2dx
+\int_{R^3}2\tau^3(u_a\cdot\nabla\widetilde{w_a}
)u_{at}\nabla\widetilde{w_a}dx\nnm\\
&\ \ \
+c\delta_T^{\frac{1}{2}}\parallel\nabla\widetilde{w_a}\parallel^2
+c\delta_T^{\frac{1}{2}}\tau
\parallel\widetilde{w_{a}}_t\parallel^2,\label{3.32}\ema
 The other
terms in the right-hand side of (\ref{3.29}) can be analyzed just
use Moser type Lemma \ref{lem2.3}, Lemma \ref{lem2.4}, the
assumptions (\ref{3.13})-(\ref{3.17}) and the Sobolev embedding
theorem, the H\"{o}lder inequality, Young's inequality. In a word,
these estimates with the above estimates (\ref{3.29})-(\ref{3.32})
will lead to \bma &\frac{d}{dt}\int_{R^3}\{\tau^3\widetilde{w_a}_t^2
   +\tau^2\widetilde{w_a}\widetilde{w_a}_t
   +\frac{1}{2}\widetilde{w_a}^2
   +\tau^3\widetilde{w_b}_t^2
   +\tau^2\widetilde{w_b}\widetilde{w_b}_t
   +\frac{1}{2}\widetilde{w_b}^2
   + \tau P_a^\prime(\rho_a^*)|\nabla\widetilde{w_a}|^2\nnm\\
  &~~~~~~~~~~~~
   +\tau P_b^\prime(\rho_b^*)|\nabla\widetilde{w_b}|^2
   +\frac{\tau \varepsilon^2}{4}(|\triangle\widetilde{w_a}|^2
   +|\triangle\widetilde{w_b}|^2)
   +\frac{\tau }{4}|\nabla\cdot\widetilde{E}|^2\}dx\nnm\\
&+\frac{d}{dt}\int_{R^3}\tau[P_a^\prime((w_a+\sqrt{\rho_a^*})^2)
-P_a^\prime(\rho_a^*)]|\nabla\widetilde{w_a}|^2dx
+\frac{d}{dt}\int_{R^3}\tau(\tau
u_a\cdot\nabla\widetilde{w_a})^2dx\nnm\\
&+\frac{d}{dt}\int_{R^3}\tau[P_b^\prime((w_b+\sqrt{\rho_b^*})^2)
-P_b^\prime(\rho_b^*)]|\nabla\widetilde{w_b}|^2dx
+\frac{d}{dt}\int_{R^3}\tau(\tau
u_b\cdot\nabla\widetilde{w_b})^2dx\nnm\\
 &+\int_{R^3}\{(2\tau -\tau^2)(\widetilde{w_a}_t^2+
   \widetilde{w_b}_t^2)+ P_a^\prime(\rho_a^*)|\nabla\widetilde{w_a}|^2+
   P_b^\prime(\rho_b^*)|\nabla\widetilde{w_b}|^2\nnm\\
&~~~
  ~~~~~~~+\frac{\varepsilon^2}{4}(|\triangle\widetilde{w_a}|^2
   +|\triangle\widetilde{w_b}|^2)
   +\frac{1}{4}|\nabla\cdot\widetilde{E}|^2+2\tau^3(u_a\cdot\nabla\widetilde{w_a}
)u_{at}\nabla\widetilde{w_a}\}dx
   \nnm\\
\leq&~ c\delta_T^{\frac{1}{2}}\parallel(\nabla w_a,\nabla
w_b)\parallel_4^2 +
c\delta_T^{\frac{1}{2}}\varepsilon^2\parallel(D^6w_a,D^6w_b)\parallel^2
+c\delta_T^{\frac{1}{2}}\parallel\nabla\cdot
E\parallel_4^2\nnm\\
&+c\delta_T^{\frac{1}{2}}\tau\parallel(D^4w_{at},D^4w_{bt})\parallel^2
+c\delta_T^{\frac{1}{2}}\parallel(\phi_a,\phi_b)\parallel_4^2,\label{3.33}
\ema
Note that the right hand side of the estimates
(\ref{3.24}),(\ref{3.28}),(\ref{3.33}) will be treated later in
terms of the estimates of $\phi_a,\phi_b$.
\par

\textbf{Step 2. The estimates for $\phi_a,\phi_b$.}\quad
Differentiating the equations (\ref{3.6}) and (\ref{3.7}) for
$\phi_a,\phi_b$ with respect to $x$, then
$\widetilde{\phi_a}=D^{\alpha}\phi_a,
\widetilde{\phi_b}=D^{\alpha}\phi_b(|\alpha|\leq4)$ will satisfy
(taking $\phi_a$ for example)
\be\tau^2\widetilde{\phi_a}_t+\widetilde{\phi_a}=D^{\alpha}
f_{a2},\label{3.34}\ee recall $f_{a2}$ in (\ref{3.10}) for $i=a.$
Taking inner product between $2\widetilde{\phi_a}$ and
(\ref{3.34}), integrating over $R^3$, we obtain
\be\tau^2\frac{d}{dt}\int_{R^3}|\widetilde{\phi_a}|^2dx
+2\int_{R^3}|\widetilde{\phi_a}|^2dx=\int_{R^3}D^{\alpha}
f_{a2}\cdot2\widetilde{\phi_a}dx,\label{3.35}\ee The terms in
right-hand side of (\ref{3.35}) can be estimated using Moser type
Lemma \ref{lem2.3}, Lemma \ref{lem2.4}, Young's inequality and the
assumptions (\ref{3.13})-(\ref{3.17}) and the inequality
$\parallel Du\parallel\leq c(\parallel\nabla\cdot
u\parallel+\parallel\nabla\times u\parallel)$ and also the
presentation of $\nabla\cdot u_a$ by $w_{at},\nabla w_a$ through
equation \ref{3.12} for $i=a$. Then we deduce
\be\tau^2\frac{d}{dt}\int_{R^3}|\widetilde{\phi_a}|^2dx
+2\int_{R^3}|\widetilde{\phi_a}|^2dx
 \leq c\delta_T^{\frac{1}{2}}\parallel \phi_a\parallel_4^2
 +c\delta_T^{\frac{1}{2}}\tau\parallel w_{at}\parallel_4^2
 +c\delta_T^{\frac{1}{2}}\parallel\nabla
 w_a\parallel_4^2\label{3.36}\ee\par

\textbf{Step 3. The closure of energy estimates.}\quad  The
assumption $\delta_T\ll1$ and the combination of the estimates
(\ref{3.24}),(\ref{3.28}) for all $|\alpha|\leq3$, and (\ref{3.33})
for all $|\alpha|=4$, (\ref{3.36}) for all $|\alpha|\leq4$ can give
us
\be\frac{d}{dt}H_1(t)+H_2(t)\leq\sum_{i=a,b}\tau\parallel
u_i(.,t)\parallel_{L^{\infty}(R^3)} \cdot\parallel
\tau^2u_{it}(.,t)\parallel_{L^{\infty}(R^3)}\cdot\parallel
D^5w_i\parallel^2\label{3.37}\ee where $H_1(t),H_2(t)$ are two
terms satisfying
$$
0<c_1 E_1(t)<H_1(t)<c_2E_1(t);~~0<c_3 E_2(t)<H_2(t)<c_4 E_2(t).
$$
for $t\in[0,T]$, and $c_1,c_2,c_3,c_4$ are positive constants
independent of $\varepsilon, \tau$, the $E_1(t),E_2(t)$ are the
terms defined in the Lemma \ref{lem3.2}. From (\ref{3.37}) we can
write
\be\frac{d}{dt}H_1(t)+H_2(t)\leq
cg(t)H_1(t),~~t\in[0,T].\label{3.38}\ee
with$$g(t)=\sum_{i=a,b}\parallel
u_i(.,t)\parallel_{L^{\infty}(R^3)} \cdot\parallel
\tau^2u_{it}(.,t)\parallel_{L^{\infty}(R^3)}$$
The assumption
(\ref{3.13}) then (\ref{3.16}) with the Gronwall inequality
applying to (\ref{3.38}) makes us know
\be H_1(t)\leq
ce^{\int_0^tg(s)ds}H_1(0)\leq ce^{c\delta_T}H_1(0)\leq
CH_1(0).\label{3.39}\ee
 for $t\in[0,T]$ provided $\delta_T\ll1$.
Integrating (\ref{3.38}) on [0,t] and using (\ref{3.39}), we derive
\be\int_0^tH_2(s)ds\leq H_1(0)+H_1(t)+C\delta_TH_1(0)\leq
C'H_1(0),\label{3.40}
\ee
The above constants $c$, $C$ and $C'$ denote the positive constant
independent of the parameters $\varepsilon>0, \tau>0$.
\par

It follows from (\ref{3.39}),(\ref{3.40}) and the equivalence
between $H_1(t)$ and $E_1(t)$, and between $H_2(t)$ and $E_2(t)$ the
conclusion stated in Lemma \ref{lem3.2}. Thus the proof of Lemma
\ref{3.2} is completed.

\subsection{The global existence and asymptotical limits}

\textsl{\textbf{The proof of global existence} (Theorem~\ref{2.6})}:
The Theorem~\ref{2.6} is a direct conclusion of the combination of
the local existence theory Lemma~\ref{local} and global a-priori
estimates Lemma~\ref{lem3.2} in terms of the variable transformation
presented above and the standard continuity argument, we omit the
details.
\par\bigskip

\noindent \textsl{\textbf{The proof of semiclassical limit}
(Theorem~\ref{semi-limit})}: Starting from Lemma \ref{3.2}, using a
continuity argument, one can easily prove the existence of the
global in-time solutions of the original problem
(\ref{1.8})-(\ref{1.11'}) with any small $\varepsilon$ and $\tau$
provided the $\Lambda_1>0$ then $\Lambda_0>0$ small enough.
\par Let
$(\psi_a^\varepsilon,\psi_b^\varepsilon,u_a^\varepsilon,
u_b^\varepsilon,E^\varepsilon)$ be the solution of
(\ref{1.8})-(\ref{1.11'}), then from Lemma \ref{3.2} and the poisson
equation (\ref{3.8}) the uniform estimates to $\varepsilon$ hold
\bma&\sum_{k=0}^{1}\parallel(\partial_t^k(\psi_a^\varepsilon-\sqrt{\rho_a^*})
,\partial_t^k(\psi_b^\varepsilon-\sqrt{\rho_b^*}))(.,t)\parallel_{5-i}^2
+\sum_{k=0}^{1}\parallel(\partial_t^ku_a^\varepsilon,
\partial_t^ku_b^\varepsilon)(.,t)\parallel_{\mathcal{H}^{5-2i}}^2\nnm\\
&+\parallel E^\varepsilon(.,t)\parallel_{\mathcal{H}^{6}}^2~\leq
~c\Lambda_0, \label{3.41}\\
&\int_0^t\{\parallel((\psi_a^\varepsilon-\sqrt{\rho_a^*})
,(\psi_b^\varepsilon-\sqrt{\rho_b^*}))(.,s)\parallel_5^2
+\parallel(\partial_t\psi_a^\varepsilon,
\partial_t\psi_b^\varepsilon)(.,s)\parallel_4^2\}ds\leq
~c\Lambda_0t\label{3.42}\\
&\int_0^t\{\sum_{k=0}^{1}\parallel(\partial_t^ku_a^\varepsilon
,\partial_t^ku_b^\varepsilon)(.,s)\parallel_{\mathcal{H}^{5-2i}}^2
+\sum_{k=0}^{1}\parallel(\partial_t^kE^\varepsilon)(.,s)
\parallel_{\mathcal{H}^{6-i}}^2\}ds\leq c\Lambda_0\label{3.43}\ema
for any $t>0$. The right-hand sides of the above inequalities are
independent of $\varepsilon$. Thus these uniform estimates and
Aubin's lemma imply the existence of subsequence denoted also by
$(\psi_a^\varepsilon,\psi_b^\varepsilon,u_a^\varepsilon,
u_b^\varepsilon,E^\varepsilon)$
such that
\bma&\psi_a^\varepsilon\rightarrow\psi_a,
~~\psi_b^\varepsilon\rightarrow\psi_b
~~in~~C(0,t;C_b^3\cap H_{loc}^{5-s}(R^3)),\label{3.44}\\
&u_a^\varepsilon\rightarrow u_a,~~u_b^\varepsilon\rightarrow u_b
~~in~~C(0,t;C_b^3\cap
\mathcal{H}_{loc}^{5-s}(R^3)),\label{3.45}\\
&E^\varepsilon\rightarrow E~~in~~C(0,t;C_b^4\cap
\mathcal{H}_{loc}^{6-s}(R^3)),\label{3.46}\ema with
$s\in(0,\frac{1}{2})$, as $\varepsilon\rightarrow0$. We also have
$$
\frac{\varepsilon^2}{2}\nabla
(\frac{\triangle\psi_i^\varepsilon}{\psi_i^\varepsilon})\rightarrow0
~~in~~L^2(0,t;H_{loc}^3(R^3))
$$
as $\varepsilon\rightarrow0$.  Thus (\ref{3.41})-(\ref{3.46}) allow
the $\varepsilon$ pass to the zero, and the limiting solutions
satisfy
$$2\psi_a\partial_t\psi_a+\nabla\cdot(\psi_a^2u_a)=0,$$
$$\tau^2\partial_t(\psi_a^2u_a)
+\tau^2\nabla(\psi_a^2u_a\otimes u_a) +\nabla
P_a({\psi_a^2})+{\psi_a^2}u_a-{\psi_a^2}E=0,$$
$$2\psi_b\partial_t\psi_b+\nabla\cdot(\psi_b^2u_b)=0,$$
$$\tau^2\partial_t(\psi_b^2u_b)
+\tau^2\nabla(\psi_b^2u_b\otimes u_b) +\nabla
P_b({\psi_b}^2)+{\psi_b}^2u_b+{\psi_b^2}E=0,$$
$$\lambda^2\nabla\cdot E=\psi_a^2-\psi_b^2
-\mathcal{C},\nabla\times E.
$$
Let $\rho_a=(\psi_a)^2 ,\rho_b=(\psi_b)^2$. It is easily to verify
that $(\rho_a,\rho_b,E)$ solves the bipolar hydrodynamic model
\ef{1.5}--\ef{1.7}. The convergence of the bipolar quantum QHD model
to bipolar hydrodynamic model is established, and the proof of the
Theorem~\ref{semi-limit} is complete.
\par
\bigskip

\noindent\textsl{\textbf{The proof of combined semiclassical and
relaxation limits} (Theorem~\ref{combine-limit})}: Since the
estimates established for the solutions in Lemma \ref{3.2} hold
uniformly for any small $\varepsilon$ and $\tau$, thus we can study
the combined limits as both $\varepsilon$ and $\tau$ tends to zero
freely.
\par Let $(\psi_a^{(\tau,\varepsilon)},\psi_b^{(\tau,\varepsilon)},
u_a^{(\tau,\varepsilon)},u_b^{(\tau,\varepsilon)},E^{(\tau,\varepsilon)})$
be the global solution derived in Theorem 2.6, by the estimates
(\ref{3.18}) in Lemma \ref{3.2}, we have the uniform estimates about
$\varepsilon$ and $\tau$ as
\bma&\parallel(\psi_a^{(\tau,\varepsilon)}-\sqrt{\rho_a^*},
\psi_b^{(\tau,\varepsilon)}-\sqrt{\rho_b^*})(.,t)\parallel_4^2
+\parallel(\tau u_a^{(\tau,\varepsilon)}, \tau
u_b^{(\tau,\varepsilon)})(.,t)\parallel_{\mathcal{H}^4}^2\leq
c\Lambda_0\label{3.47}\\
&\parallel(\tau\partial_t\psi_a^{(\tau,\varepsilon)}
,\tau\partial_t\psi_b^{(\tau,\varepsilon)})(.,t)\parallel_3^2
+\parallel
E^{(\tau,\varepsilon)}(.,t)\parallel_{\mathcal{H}^5}^2\leq
c\Lambda_0,\label{3.48}\ema
 and
\bma&
\int_0^t(\parallel(\psi_a^{(\tau,\varepsilon)}-\sqrt{\rho_a^*},
\psi_b^{(\tau,\varepsilon)}-\sqrt{\rho_b^*})(.,s)\parallel_5^2
+\parallel(\partial_t\psi_a^{(\tau,\varepsilon)}
,\partial_t\psi_b^{(\tau,\varepsilon)})(.,s)\parallel_3^2)ds\nnm\\
\leq& c\Lambda_0t,\label{3.49}\\
&\int_0^t(\parallel( u_a^{(\tau,\varepsilon)},
u_b^{(\tau,\varepsilon)})(.,s)\parallel_{\mathcal{H}^4}^2+\parallel
E^{(\tau,\varepsilon)}(.,s)\parallel_{\mathcal{H}^5}^2)ds\leq
c\Lambda_0,\label{3.50}\ema
 for any $t>0$.
\par Also use Aubin's lemma with the above uniform estimates, we
can get the subsequence(not relabelled) and functions denoted also
by $\psi_a,\psi_b,u_a,u_b,E$ such that as $\varepsilon,\tau
\rightarrow0$ \bma&\psi_a^{(\tau,\varepsilon)}\rightarrow\psi_a,
~\psi_b^{(\tau,\varepsilon)}\rightarrow\psi_b~~in~~C(0,t;C_b^2\cap
H_{loc}^{4-s}(R^3)),\label{3.51}\\
&u_a^{(\tau,\varepsilon)}\rightharpoonup
u_a,~u_b^{(\tau,\varepsilon)}\rightharpoonup u_b
~~weakly~~in~~L^2(0,t; \mathcal{H}^{4}(R^3)),\label{3.52}\\
&E^{(\tau,\varepsilon)}\rightarrow E~~in~~C(0,t;C_b^3\cap
\mathcal{H}_{loc}^{5-s}(R^3)),\label{3.53}\ema
 for any $t>0$ and
$s\in(0,\frac{1}{2}).$
\par From (\ref{3.47}),(\ref{3.48}) we know $\psi_a,\psi_b$
are positive in $(0,t)\times R^3$, and also
\be\tau^2|u_a^{(\tau,\varepsilon)}|^2\rightarrow0,
~~\tau^2|u_b^{(\tau,\varepsilon)}|^2\rightarrow0
~~in~~L^1(0,t;W_{loc}^{3,3}(R^3)),
~as~~\tau,\varepsilon\rightarrow0.\label{3.54}\ee
 Thus the above
converging results allow the solutions to pass to the limit
$\tau,\varepsilon\rightarrow0$ from the bipolar QHD model to the
bipolar drift-diffusion(DD) model:
$$2\psi_a\partial_t\psi_a-\nabla\cdot[\nabla P_a((\psi_a)^2)-(\psi_a)^2E]=0,$$
$$2\psi_b\partial_t\psi_b-\nabla\cdot[\nabla P_b((\psi_b)^2)+(\psi_b)^2E]=0,$$
$$
\lambda^2\nabla\cdot E=(\psi_a)^2-(\psi_b)^2-\mathcal{C},
 \nabla\times E=0,
$$
 which is equivalent to the bipolar DD model (\ref{1.13})-(\ref{1.14})
in Section 1 for strong solution. Namely, $(\rho_a=(\psi_a)^2
,\rho_b=(\psi_b)^2,E)$ solves the bipolar Drift-Diffusion model
(\ref{1.13})-(\ref{1.14}). The proof of Theorem 2.7 is completed
now.

\bigskip

\noindent\textbf{Acknowledgements:} The authors acknowledge the
partial support by the National Science Foundation of China
(No.10571102), the Key Research Project on Science and Technology of
the Ministry of Education of China (No.104072), the grant- NNSFC
(No.10431060), Beijing Nova program, and the Re Shi Bu Ke Ji Ze You
program.

\end{document}